# A new parsimonious method for classifying Cancer Tissue-of-Origin Based on DNA Methylation 450K data

**Shen Jia[1#], Yulin Zhang[2#], Yiming Mao[1], Jiawei Gao[1], Yixuan Chen[1],**

**Yuxuan Jiang[1], Haochen Luo[3], Kebo Lv[4*], Jionglong Su[5*]**

[1] *Department of Mathematics, University of Liverpool, Liverpool, UK*

[2] *College of Mathematics and Systems Science, Shandong University of Science and Technology, Qingdao, China*

[3] *Department of Computer Science, University of Liverpool, UK*

[4]*School of Mathematical Sciences, Ocean University of China, Qingdao, China*

[5] *School of AI and Advanced Computing, XJTLU Entrepreneur College(Taicang), Xi'an Jiaotong-Liverpool University, Suzhou, China*

*# These authors contribute equally to the work*

**\*Corresponding author: Kebo Lv (e-mail: kewave@ouc.edu.cn),**

**Jionglong Su (email: Jionglong.Su@xjtlu.edu.cn)**

# *Abstract*

DNA methylation is a well-studied genetic modification that regulates gene transcription of Eukaryotes. Its alternations have been recognized as a significant component of cancer development. In this study, we use the DNA methylation 450k data from The Cancer Genome Atlas to evaluate the efficacy of DNA methylation data on cancer classification for 30 cancer types. We propose a new method for gene selection in high dimensional data(over 450 thousand). Variance filtering is first introduced for dimension reduction and Recursive feature elimination (RFE) is then used for feature selection. We address the problem of selecting a small subsets of genes from large number of methylated sites, and our parsimonious model is demonstrated to be efficient, achieving an accuracy over 91%, outperforming other studies which use DNA micro-arrays and RNA-seq Data . The performance of 20 models, which are based on 4 estimators (Random Forest, Decision Tree, Extra Tree and Support Vector Machine) and 5 classifiers (k-Nearest Neighbours, Support Vector Machine, XGboost, Light GBM and Multi-Layer Perceptron), is compared and robustness of the RFE algorithm is examined. Results suggest that the combined model of extra tree plus catboost classifier offers the best performance in cancer identification, with an overall validation accuracy of 91% , 92.3%, 93.3% and 93.5% for 20, 30, 40 and 50 features respectively. The biological functions in cancer development of 50 selected genes is also explored through enrichment analysis *and the results show that 12 out of 16 of our top features have already been identified to be specific with cancer and we also propose some more genes to be tested for future studies. Therefore, our method may be utilzed as an auxiliary diagnostic method to determine the actual clinicopathological status of a specific cancer.*

# 1 Introduction

## 1.1 DNA methylation and cancer identification

DNA methylation is an epigenetic modification of the genome that is crucial for the normal regulation of gene transcription [1]. It is a biological process involving the transfer of a methyl group onto the C5 position of the cytosine to form 5-methylcytosine [2]. In normal cells, this modification results in different interaction properties assuring the proper regulation of gene expression and of gene silencing [3]. In contrast, malignant cells show major disruptions in their DNA methylation patterns. Alternations of DNA methylation consist of hypomethylation and hypermethelation have been recognized as an important component of cancer development. (Hypomethylation is usually linked to activation of oncogenes, whereas hypermethylation of the CpG islands is associated with silence of cancer suppressor gene [4].) With the DNA methylation profile, it is expected that classification of cancer can be achieved based on different levels of DNA methylation, which may be helpful in tumor diagnosis and drug development [1,5]. Generally, it is known that cancer is a group of diseases that are driven by progressive genetic abnormalities that include mutations in tumor-suppressor genes and oncogenes, as well as chromosomal abnormalities [6]. While the role of genetic mutations has been highlighted by many researchers [7-11], epigenetic alternations such as DNA methylation have also been found highly related to cancer development [12].

Traditionally, cancer identification is carried out through medical tests such as lab tests, imaging tests and biopsy [13]. Previous work of Masilamani et.al [14] suggests that lab tests such as blood tests for tumor markers have great potential for early detection of cancer. However, as pointed out by National Cancer Institutes (NCI), abnormal lab results are not a sure sign of cancer and most of them may only be helpful in assessment of cancer risk rather than disgnosis [15]. Likewise, imaging tests are helpful in the early detection of cancer. They can help locate and find out the stage of the tumor. However, due to some similar appearance between cancer and other types of diseases, they are not suitable for making a final diagnosis. Additionally, since it takes millions of cells to make a tumor big enough to appear on an imaging test, imaging test may only be helpful in detecting large groups of cancer cells [16]. In terms of biopsies, although they provide the most accurate diagnosis of cancer, a sample is needed to run such tests. Usually, the biopsy sample is obtained with a needle, endoscopy or even a surgery and these may cause great pain to patients [13].

Considering the limitations of current techniques, we hope to provide additional tools for cancer indentification and classification from a molecular level using sequencing DNA methylation data. The genetic analysis of methylated DNA data may also provide new insights into drug development.

### 1.2 Sequencing data improves cancer classification

Accurately identifying specific cancer holds great promise for further treatment selection and prediction of prognosis in cancer [12]. However, due to the limitations of surgical tolerance in patients, it might be difficult for the surgeons to use complex anatomy to obtain cancer diagnosis in high accuracy [17]. Sometimes, even when the tumors are successfully obtained, similar histopathological appearance of different tumors may still lead to misdiagnosis and consequently treatment may not achieve effective therapeutic effect [18]. The uncertainty of this kind of prognosis propels the development of other diagnostic aprroaches with higher certainty. With the improvement of the Next-Generation Sequencing (NGS) technology, more and more sequencing data e.g., DNA-seq, RNA-seq are available to researchers which require advanced methods to mine these data and provide a more general and reliable information for clinical diagnosis [19] through specific biological insights. A great number of research have been carried out for classifying cancer types and subtypes using sequencing data. For instance, Li(2017) and his team, who used the RNA-seq Data to extract 20 features in the classification of 33 different types of tumors with the mean of ML, with the final accuracy rate exceeded 90% [20]. Similar thoughts could be seen from Zhang (2018)'s project, where they proposed a prognosis model based on Bayesian network classification of breast tumor with 888 features extracted from DNA methylation Data [21]. Raweh (2018) and his team used TCGA data sets and RnRead software to extract the 512 useful feature from the raw data, then utilized the method of fast Fourier transform algorithm for seven types of cancer respectively extracted less than 10 corresponding characteristics and finally used different classifiers (such as random forest) and support vector machine to get an overall accuracy which was above 97% [22]. However, despite the fact that many techniques have been applied to analyze gene expression data, it has been demonstrated by previous studies that some common machine learning methods such as Support Vector Machine may not perform well due to the high dimensionality and small sample size of Gene Expression Profiles (GEP) dataset [23]. Thus, the design of dimensional reduction method to drastically reduce the number of features in sequencing data would be the key to acquire a model in high accuracy.

### 1.3 Our work

In this study, we use the DNA methylation 450k data from The Cancer Genome Atlas to evaluate the efficacy of DNA methylation data on cancer classification for 30 cancer types. Since the dimension of the data is extremely high(over 450 thousand ), we first use the variance selection method to reduce dimension and then utilize the recursive feature elimination algorithm to extract the features. Enrichment analysis is then carried out to explore the biological meaning of the selected genes.

#### 1.3.1 Motivation

The motivation of this paper is two-fold:
(i) First, inspired by Darst who acknowledged the positive impact of Random Forest-Recursive Feature Elmination algorithm in extracting features from smaller data sets, our team apply the algorithm into high dimensional data[24].

(ii) Second, the 450k methylation value provides a genome-wide quantitative representation of DNA methylation value in a convenient format, which makes the data preprocessing process for analysts easier and more feasible and thus arouses our interest to explore its correlation with cancer[25].

### 1.3.2 Novelty

The novelty of this paper is three-fold:

(i) First, many previous works[20,21] dived into the area where the sequencing data was used to diagnosing cancer type, but to the best of our knowledge, none of them focused on the DNA methylatyion 450k data to carry out the cancer classification.

(ii) Second, instead of building a model with hundreds of identifiers like Zhang (2018) and Raweh (2017)did, our team focus on the feasibility to construct a parsimonious model with relatively fewer number of identifers (within 50 features).

(iii) Third, we prepose a novel method which includes variance selection, recursive feature elimination and boosting classifiers in extracting features from high dimensional data.

### 1.3.3 Key contributions

The key contributions of this paper are two-fold:

First, we have demonstrated the efficiency of our parsimonious model, with an overall accuracy of 91% , 92.3%, 93.3% and 93.5% for 20, 30, 40 and 50 features respectively, achieving the same level of accuracies of over 90% as other studies which use DNA micro-arrays and RNA-seq Data.

Second, 12 out of 16 top features in our studies have been already shown to play a part in cancer development, and we thus recommend some more genes for further testing.

This paper is divided into four sections as follows:

1) Section I is the introduction. It briefly describes the background of the cancer identification and the significance of DNA methylation for cancer identification .
2) Section II contains methodology. This part mainly introduces the source of the dataset and how our work is conducted in the two phases: dimensionality reduction and feature selection.
3) Section III Results and Discussions. This chapter presents the result of our alogirthm and compared the performance of models with different number of features and classifiers. At the same time, it also includes enrichment analysis and gene annotations to explore the biology meaning of our works
4) Section IV contains the conclusion of this article and the work we expect to do in the future.

## 2  Methodology

### 2.1 Dataset

The Cancer Genome Atlas (TCGA) is a landmark project led by the National Cancer Institute (NCI) and National Human Genome Research Institute (NHGRI) that aims to catalogue major cancer-causing genetic mutations [26]. The TCGA dataset, molecularly characterized over 20,000 primary cancer and matched normal samples spanning over 30 cancer types, is publicly available [27, 28].

This study focuses on the DNA methylation data from the TCGA dataset provided by UCSC Xena. DNA methylation profile was obtained using the Illumina Infinium HumanMethylation450 platform which includes more than 450 thousand CpG site probes(HumanMethylation450), which provides quantitative methylation measurement at the single-CpG-site level. Featuring more than 450 thousand methylation sites, the Infinium HumanMethylation450 BeadChip offers a combination of comprehensive, expert-selected coverage and high throughput at a low price, thus enabling cost-effective DNA methylation analysis for a variety of applications [29]. Due to its comprehensive genome-wide coverage as well as incomparable cost-effectiveness, the HumanMethylation450 datasets have become a popular choice for the study of epigenetic changes in many diseases processes [29-31]. Saumya used DNA Methylation 450K data to reveal the peripheral blood differential methylation in male infertility and Bonnie utilized the same data to make the MGMT mrthylation assessment in glioblastoma[32,33].

In this study, the methylation 450kdata we use derived from UCSC Xena has already gone through quality control such as filtration of low quality probes as well as normalization process and DNA methylation level is measured in beta value defined as the ratio of the methylated probe intensity and the overall intensity (sum of methylated and unmethylated probe intensities) [34].

$$\beta_n = \frac{\max(Meth_n, 0)}{\max(Meth_n, 0) + \max(Unmeth_n, 0) + \alpha}$$

which is a continuous variable between 0 and 1. Higher beta value represents higher level of DNA methylation whereas lower beta value represents lower level of DNA methylation.

## 2.2 Dimension Reduction

The study focuses on 30 DNA methylation datasets extracted from TCGA (listed in Table 1). The original data consists of over 450 thousand identifiers representing the ID number of the probes and 9743 samples in total, with a number of features containing missing value due to low quality of certain Methylation450 BeadChip probes.

Inspired by the idea of neural network, the feature selection from the original 480,000 CpG sites is implemented in two-phases (*Figure 1*). we propose a 'hidden-layer' (Medium Layer)size features which is the output of the variance selection

(Dimension Reduction) as well as the input of the latter RFE[35] process (Feature Selection).

In the first step, by variance selection method (Dimension Reduction), we reduce the number 450,000 to 2000 features.

In the second step, the first phase's output becomes the input of the RFE process (Feature Selection) and we extract 20 features from the 2000 CpG sites, which will be later discussed in 2.3. The reason we construct a 'hidden-layer' here is that though RFE can provide precise solution, the time overhead is a defect when the input features are enormous[35]. In addition, if we direcly apply RFE algorithm to 480,000 featrues, the algorithm's stability is questioned since redundant and noisy genes are not removed at first as Tang proved [36].

Before the dimensionality reduction and feature selection, features with missing values in the samples were removed and then 30 datasets were merged into one overall dataset.

After that, Dimension reduction was conducted based on variance between and within different cancer groups . Afshar(2020) demonstrated the effectiveness of removing features with low variance so as to select important features in high dimensional data [37]and Model(2001) showed that genes with large variance are more important features since they explain most of the total variances[38]. Therefore, we combined their thoughts and applied the variance selection in this study.

The following are the formulas of the within groups variance $s_{n,j}^2$ and between groups variance $S_j^2$ respectively. In  formula (1), $x_{i,j}^{(n)}$ represents the value in row(Sample) *i* and column(feature) *j* of nth class. In addition, $\bar{x}_j^{(n)}$ represents the mean of all the values in column(feature) *j* of nth class. $N_n$ represent the number of sample each cancer class has. Similarly, in the latter, $x_{i,j}$ represents the value in row(Sample) *i* and column(feature) *j*., $\bar{x}_j$ represents the mean of all the values in column(feature) *j*. $N$ represents the number of total samples.

$$s_{n,j}^2 = \sum_{i=1}^{N_n} \frac{\left(x_{i,j}^{(n)} - \bar{x}_j^{(n)}\right)^2}{N_n - 1} \quad (1)$$

$$S_j^2 = \sum_{i=1}^{N} \frac{(x_{i,j} - \bar{x}_j)^2}{N - 1} \quad (2)$$

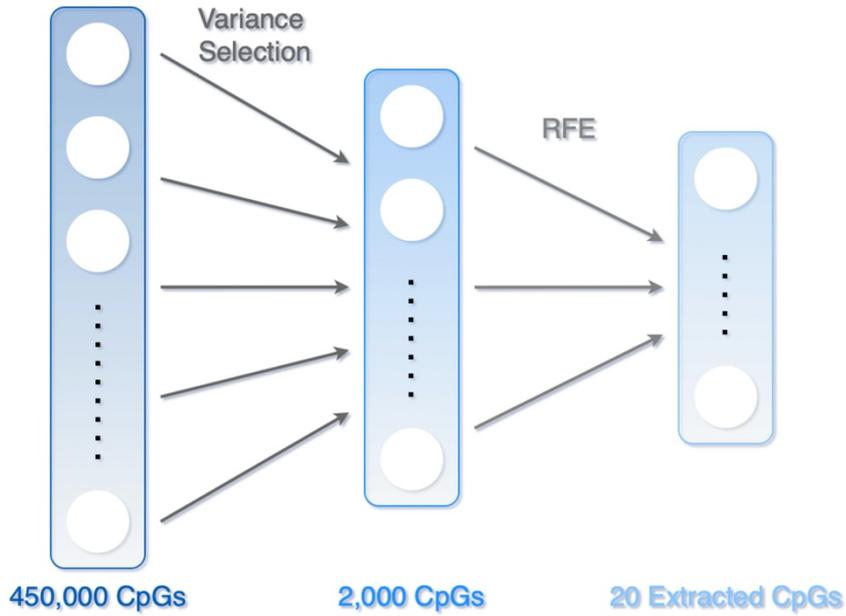

**FIGURE 1** | Dimension Reduction and Feature Selection

*Figure 1 Dimension Reduction and Feature Selection*

### 2.3 Features selection

In this part, we mainly utilize the feature selection algorithm recursive feature elimination method (RFE) method[35] to obtain the most important features from a total of 2000 identifiers after the first step. As the work of Guyon et.al [35] illustrates, this algorithm achieve high accuracy in genetic diagnosis and gene selection. More specifically, RFE is a method for feature selection which selects features by recursively obtaining increasingly smaller sets of features until gets the required number of features. An estimator, namely a classifier gives the impotrance of the features, is an essentail part of this algorithm, which computes the significance of the features. In each loop, it ranks the significance of the features and eliminates the worst feature. Here, we choose accuracy as the indicator for the RFE and mainly tune the number of chosen features to compare the results.

Since the RFE method is estimator bounded, the choice of it can be vital. According to Deng(2012), tree-based classifiers are frequently chosen as the estimator since they can provide variable significance scores and perform relatively strong[39]. Similar points can be seen in Touw(2013) 's paper, which demonstrates the significance of random forest classifier to the bioinformatics area[40]. Therefore, random forest are chosen in this paper as the estimator and we also proposed its similar algorithm decision tree [41] and extra tree [42] to be tested.

On the other hand, Moon(2016) demonstrated that the L1-norm SVM performed well as the feature selection estimators when dealing with the biomedical data[43]. Reasons are that the figure for the biostatisticians tend to be centered in an interval which can be hard to distinguish while L1-based classifiers are able to make them sparse and easier for detect. As a result, since the data we have are the decimals between 0.0 and 1.0(mainly 0.1 to 0.9 according to the information on https://xenabrowser.net) , we follow Moon and his team to use the L1-norm SVM as the estimator

Therefore, we aim to use these four estimators to get the scores for each feature and then select the top several to feed into the classifiers. To mention above, fivedifferent classifiers are used in this paper. Following Sahu(2017) 's research[44] on the DNA microarray data, Support Vector machines (SVM) [45], k-Nearest Neighbors (KNN) [46] and    Multilayer Perceptron (MLP) 2.3[47] are chosen. Besides, another two ensemble classifiers LightGBM [48] and Xgboost[49] are also used.

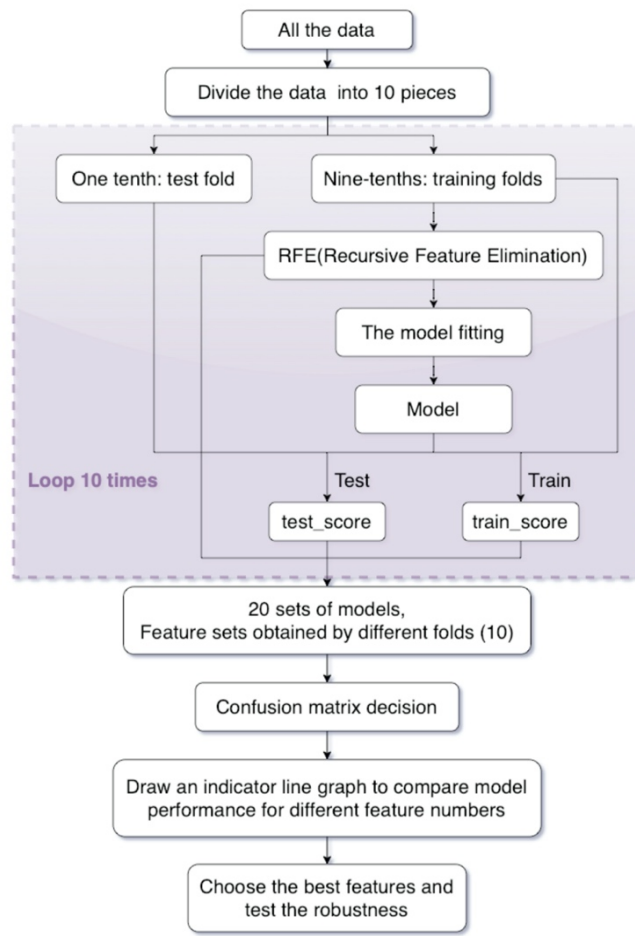

**FIGURE 2** | Flow chart of method

The feature set filtered by variance was first divided into 10 parts for 10 folds cross validation(Figure 2). In each iteration, keep one fold of the data as the validation set and use the other 9 parts as the training set. Then we use the RFE method to fit on the training set and achieve the selected features. Thus, a new subset of the data, written by using the selected CpG sites as features, enters into the classifier to be evaluated. Here, the key point is that we need to check the performance of the selected features as well as choosing the best group (features plus the classifier) as the final model.

Since there are four estimators as well as five various classifiers in the experiment, a total number of 20 groups are made and each group contains 10 iterations since a ten folds cross-validation is contained in each group. As shown in the graph(*Figure 3*), the training set and test set were tested respectively with the models (the motion for testing the training part is to check whether the model is overfitting), and the accuracy was recorded at intervals of 10 features from 20 to 50. In this way, two lists each containing    train score and test score respectively are obtained. Therefore, for each of the 20 models, we obtained accuracy for each folds on the validation set and averaged them to get one mean set accuracy for the different number of features as well as calculated the minimum and maximu value of the 10 groups.

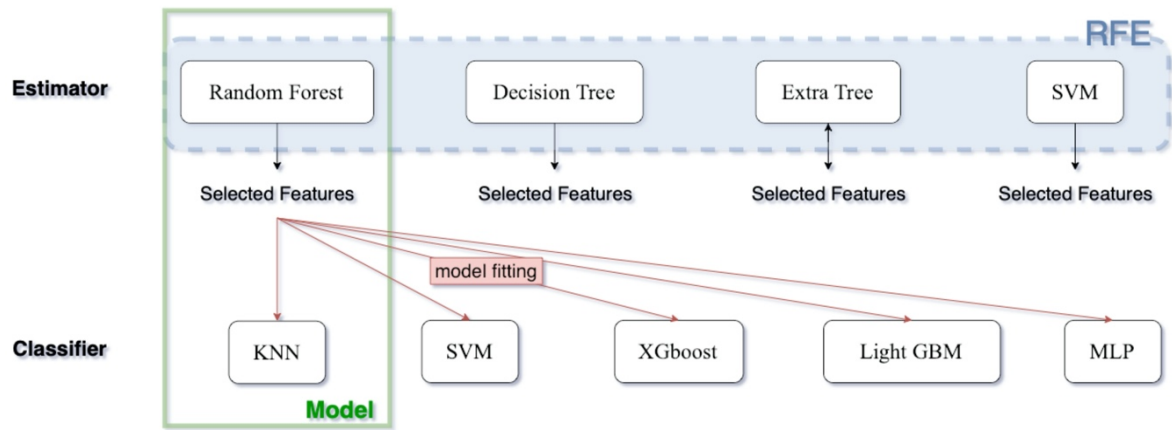

*Figure 3 Combination of estimator and classifier*

## 3    Results and Discussions

In this section, we apply our feature selection method to extract important features from the dataset and then analyze the results. The biological meaning of the figure will also be included.

### 3 .1    Extra Tree as the Estimator

In the methodology part, we have mentioned that an overall 20 models are used to look for the best estimator for the RFE algorithm, and the results are as followed. We did the 10 fold cross validation to each of these combined models and same to other paper[50], the accuracy here is based on its performance on the test set. The number of features we first selected is 20 as our novelty here is to look for models with least features but also keep a relatively decent accuracy.

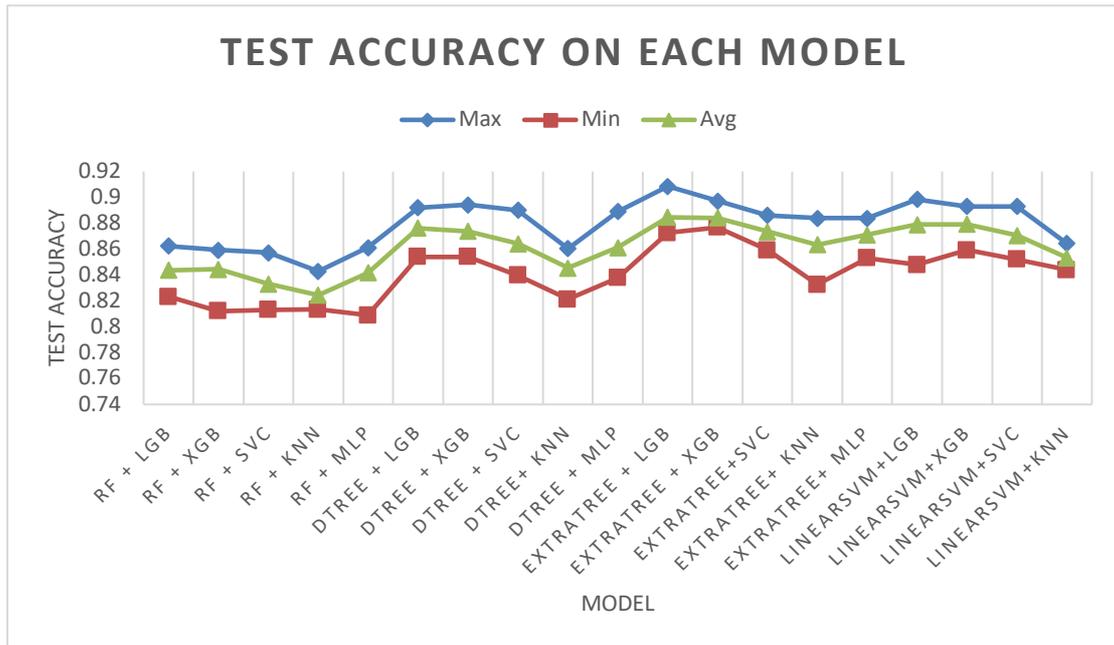

*Figure 4 test accuracy on each model*

From *Figure 4*, extra tree algorithm is selected as the estimator of the RFE algorithm since the model with the extra tree have higher 10 cross validation accuracy compared to others. Also, since boosting classifier as lightgbm and xgboost performance well with our model, we additionally test the score of the Catboost classifier [51].

### 3.2 Model performance for 20 points

We first make our experiment on 20 points since it is the proposed lower bound for our feature numbers. According to 1.1, we obtain the 20 features as below.

| Cg08875705 | Cg22727783 | Cg13668207 | Cg05738240 | Cg24402880 |
| Cg17853216 | Cg12635790 | Cg03868944 | Cg13709271 | Cg14528056 |
| Cg23473955 | Cg17295878 | Cg17198308 | Cg00518941 | Cg10706013 |
| Cg08552167 | Cg07823492 | Cg18368125 | Cg27009703 | Cg02125316 |

*Table 1 20 selected CpGs*

In the research, the sklearn metrics to access the performance of the model includes true positive, true negative, false positive and false negative. Several criteria such as recall score, precision score, f1 score and the overall accuracy will be used as the main indicator for the paper. The task is a multi-class classification, so we choose the parameter 'weighted' to be the key parameter for the indicators since the samples of each class in the dataset is not totally the same. In the following formulas, $TP_i$ represents the number of sample that the truth is class i and the test predicts class i ; $TN_i$ represents the number of sample that the truth is not class i and the test predicts not class I ; $FT_i$ represents the number of sample that the truth is class i but test predicts not class i ; $FP_i$ represents the number of sample that the truth is not class i but test predicts class i[52].

$$Overall\_Accuracy = \frac{\sum_{i=1}^{30}(TP_i + TN_i)}{\sum_{i=1}^{30}(TP_i + FP_i + TN_i + FN_i)}$$

$$Precision = \frac{\sum_{i=1}^{30} TP_i}{\sum_{i=1}^{30}(TP_i + FP_i)}$$

$$Recall = \frac{\sum_{i=1}^{30} TP_i}{\sum_{i=1}^{30}(TP_i + FN_i)}$$

$$F1_{score} = \frac{2 \times Precision \times Recall}{Precision + Recall}$$

The overall accuracy is the proportion of correct predictions among the total number of case examined. Precision is the Agreement of the data class labels with those of a classifiers if calculated from sums of per-text decisions. Recall is the Effectiveness of a classifier to identify class labels if calculated from sums of per-text decisions. . The $F_1$ score is the harmonic mean of the precision and recall. [52-54]

And then we get the model performance of the three classifiers with only 20 features based on these indicators.

|         | $Overall\_Accuracy$ | $Precision$ | $Recall$ | $F1_{score}$ |
|---------|---------------------|-------------|----------|--------------|
| **CATBOOST** | **0.91009**    | **0.91866** | **0.91009** | **0.91320** |
| LGB     | 0.89305             | 0.90163     | 0.89305  | 0.89603      |
| XGB     | 0.89418             | 0.90156     | 0.89418  | 0.89675      |

*Table 2 Compared performance of three classifiers*

Thus, the combined model of extra tree plus catboost classifier offer best performance on the dataset and we use this combination to demonstrate the confusion metrics (*Figure 5*).

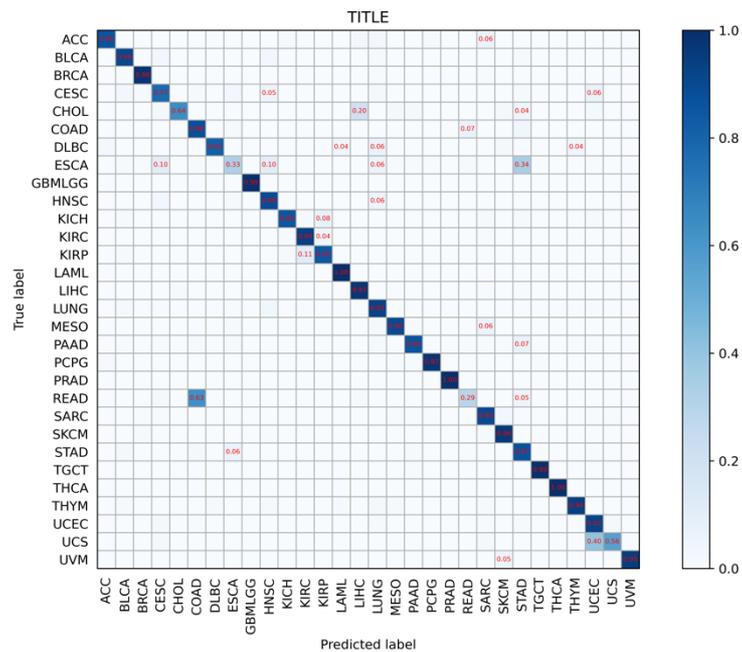

*Figure 5 Confusion metrics for 20 points*

The element in the diagonal of the matrix represent the sample which can be predicted correctly from the model, and all the other element in the confusion metrics are the wrong predicted samples[55].

From the results, 17 out of 30 classes have the accuracy over 90%, 25 outof 30 classes are able to predict accurately over 80%, however, there still exists some classes which difficult to distinguish between. 34 percent of the ESCA sample are wrongly predicted as STAD, 63 percent of the READ data inaccurately predicted as COAD and 40 percent of the UCS samples falsely predicted as the UCEC.

The imbalanced data set can be one of the factor. Clearly, according to Appendix 1, the sample number we have for UCS is only 57, while UCEC has a relatively larger class at the number of 478, so it may reasonably predicted as the other class which own relatively larger number of data due to the quality of data.

On the other hand, similar to Li(2017)'s work, though they achieve an overall high accuracy on the test set, they still fail to classify three tumor types, such as READ(rectum adenocarcinoma) and COAD(colon adenocarcinoma).

Our team here again showed the difficulty of classifying between READ and COAD class and we may test whether increasing the features can solve the problem in 3.3

**3.3 Compared model performance for different feature numbers**

Similar to 3.2, we directly use the extra tree as the estimator for the rfe to extract features and use the catboost classifier as the predicting model. The indicators we use here are the same as above and get the result as below(*Figure 6*)

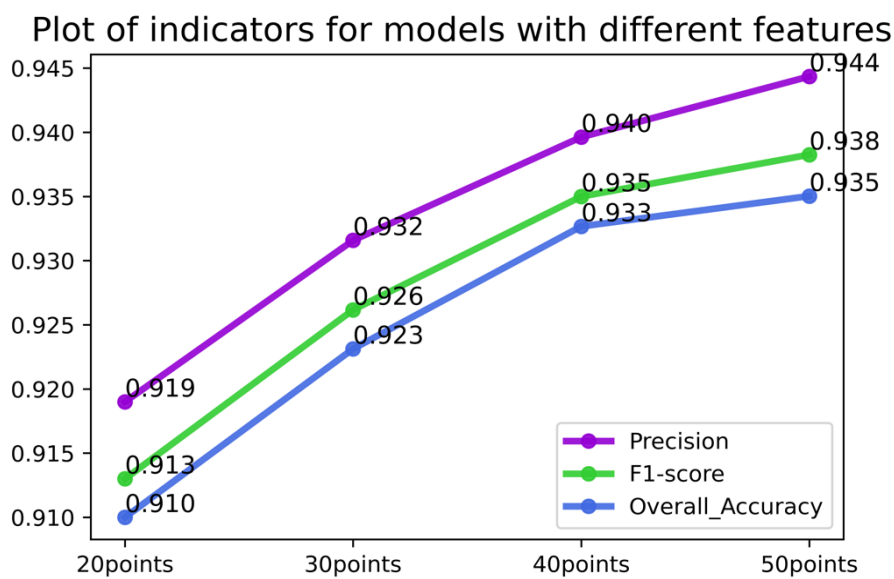

*Figure 6*

From figure 6 , the performance of the model increases continuously until reaching 40 points, the slope here starts to decrease. Since our aim is to look for the solution with best efficiency (the lowest number of features with high accuracy), we prefer the solution with 40 features. We then analyze the confusion metrics of the models as below (Figure 7, Figure 8 and Figure 9).

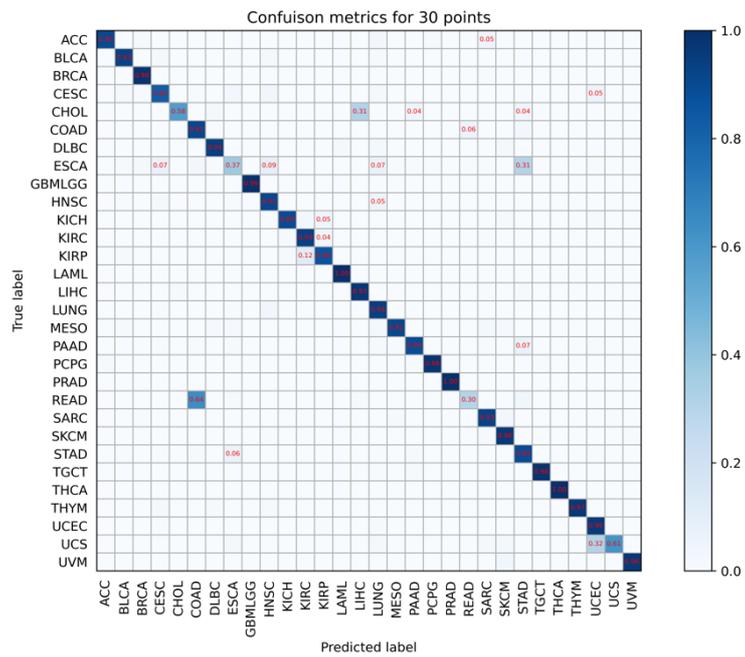

*Figure 7 Confusion metrics for 30 features*

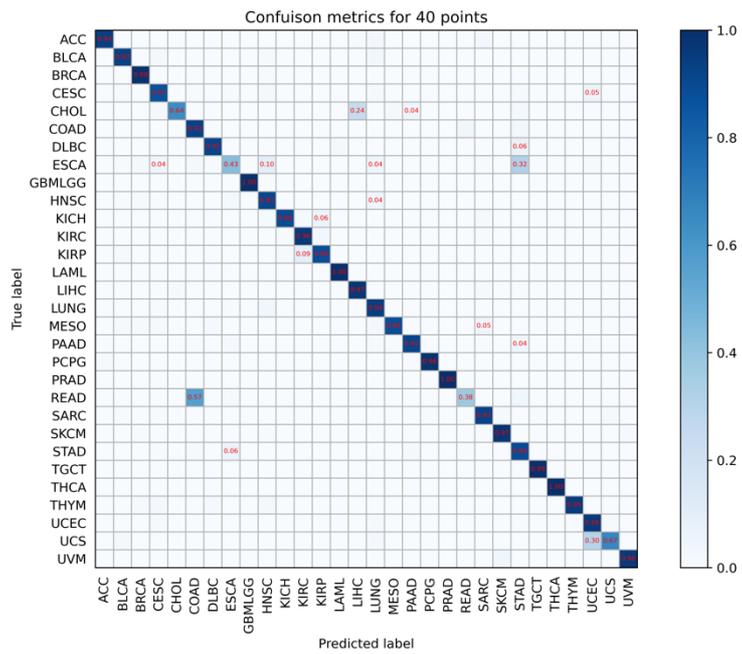

*Figure 8 Confusion metrics for 40 features*

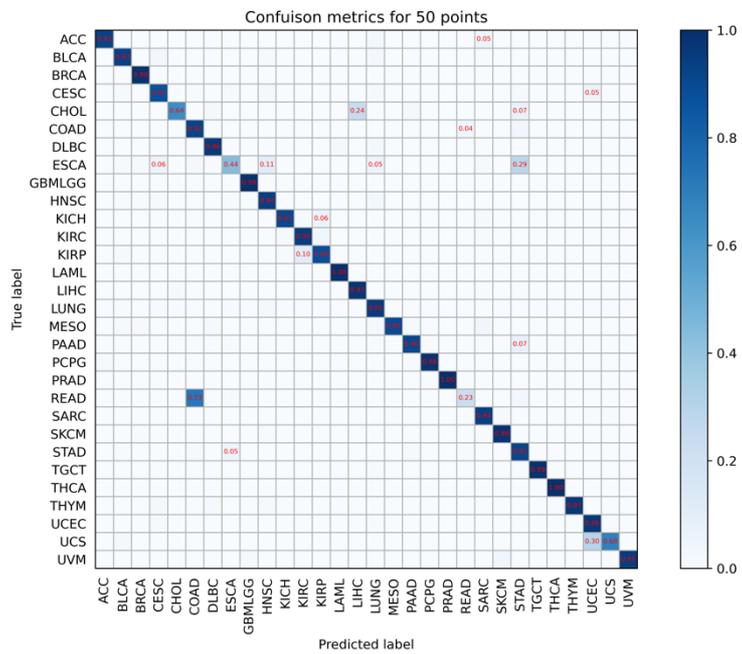

*Figure 9 Confusion metrics for 50 features*

From those figures, the ambiguity between COAD and READ still exists and not better with the increasing number of features.

The UCS class has an obviously higher accuracy compared to 20 points and so does the ESCA class, so we point out the possibility that these two classes can achieve a decent overall cross-validation accuracy with enough number of features. However, since our paper mainly focuses on classifying these cancer types based on tiny number of features, we will not include further discussion here. Meanwhile, 21 of 30 class with 40 features can achieve accuracy over 0.9 compared to 17 for the 20 features and. In a nutshell, our solution using only 40 identifiers is able to predict the 30 cancer types with 93.3% overall accuracy using the data of DNA methylation 450 values.

**3.4 Test robustness of the model**

This section works as an inspection for the model, we need to examine the convergence of the RFE algorithm.

**3.4.1   RFE Inspection**

First, our team examine the intersection of the features. Since RFE algorithm is a backward stepwise selection method which means that the solutions of fewer feature should better be included in the larger feature set. Then we calculate the exact intersection set as followed.

| Overlap | 30 points | 40 points | 50 points |
|---|---|---|---|
| For 20 points | 20 | 17 | 18 |
| For 30 points | / | 25 | 27 |
| For 40 points | / | / | 35 |

*Table 3 elements of intersection set*

Therefore, the convergence of our model is totally good since the inclusion rate is high. To be mentioned, since the 2000 filtered identifiers need to be reduced to 20, the first twenty selected features should be most robust, then the second twenty features (20-40). So it is reasonable that the overlapped features of 20 points with the 50 points are greater than the 20 points with the 40 points.

Second, we dive into the performance of the intersection points.

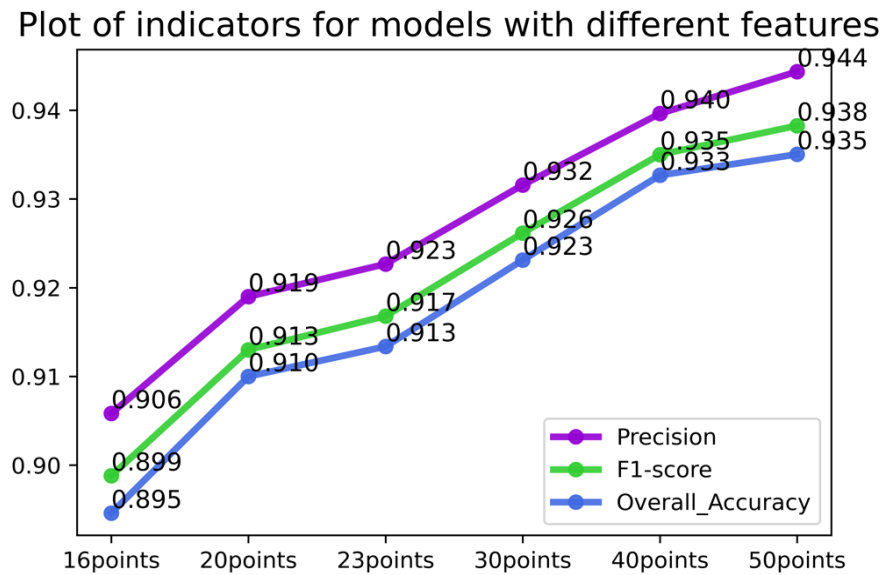

*Figure10Plot of indicators for models consisting of different number of CpGs*

From above, it shows that the 16 points solution (the intersection of the 20 points and the 30 points) and the 23 points solution (the intersection of the 30, 40, 50 points) do follow the trend. One point to be noticed is that the indicators decrease for features fewer than 20, which demonstrate that 20 features can be a good starting point for the experiment.

### 3.5 Genes and biological meaning

#### 3.5.1 Enrichment Analysis

After extracting the identifiers(CpGs) from the data, we map back to find the related genes(Prefix 2) and enrichment analysis has been carried out with DisGeNET[56] and GO Biological Processes [57]by using the *metascape*[58]. Terms with a p-value[59] < 0.01, a minimum count of 3, and an enrichment factor > 1.5 (the enrichment factor is the ratio between the observed counts and the counts expected by chance) are collected and grouped into clusters based on their membership similarities. More specifically, p-values are calculated based on the accumulative hypergeometric distribution, and q-values are calculated using the Banjamini-Hochberg procedure[60] to account for multiple testings. The results below is the summary of enrichment analysis in DisGeNET

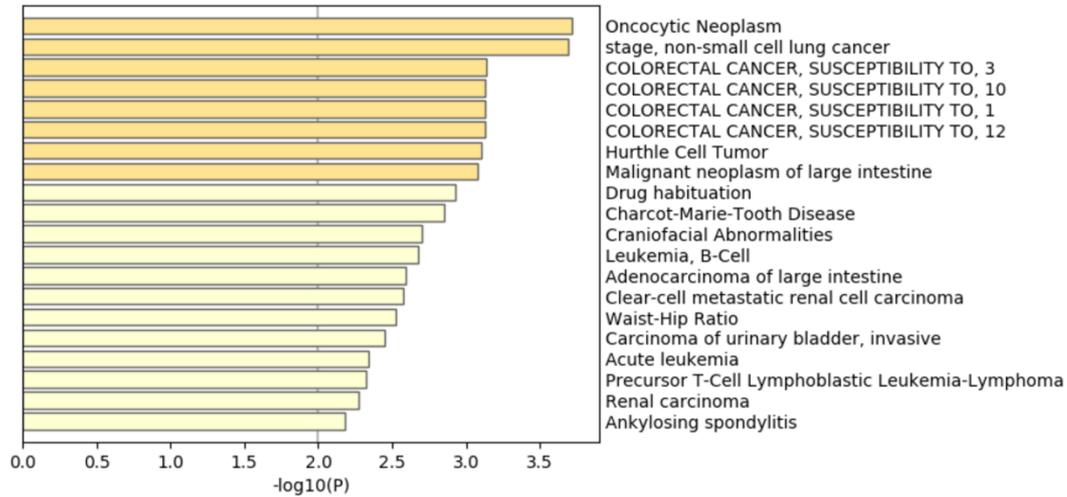

*Figure 11 Summary of enrichment analysis in DisGeNET*

According to Piñero et al (2019), DisGeNET collected genes and variants associated with human diseases. Therefore, our purpose is to test whether the features we extract by the numerical algorithm have its biological meaning. To be mentioned, since the indicator here is $-log(P)$, which means that a larger figure leads to a greater significance. Therefore, we identified the first eight diseases in the graph which all turns out to be cancer or related malignant(Oncocytic Neoplasm, Lung Cancer, Colorectal cancer, Thyroid Gland Hurthle (Oncocytic) Cell Neoplasm Hurthle and Malignant Colorectal Neoplasm[61-64]).

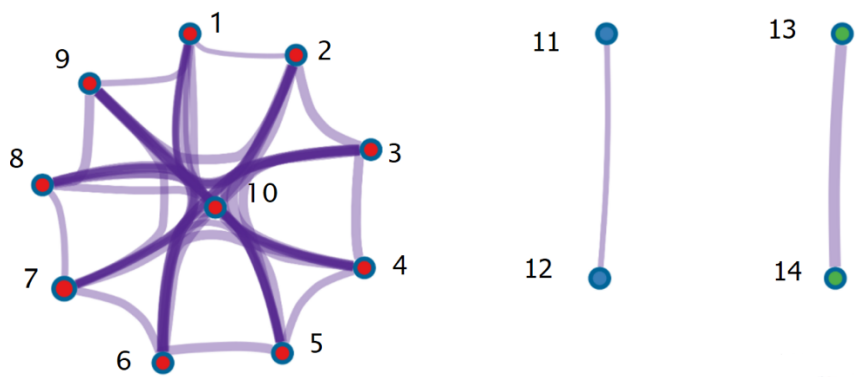

*Figure 12 Network of enriched terms colored by cluster ID*

Next, we identified all statistically enriched terms accumulative hypergeometric p-values and enrichment factors were calculated and used for filtering. Remaining significant terms were then hierarchically clustered into a tree based on Kappa-statistical similarities among their gene memberships. Then 0.3 kappa score[65] was applied as the threshold to cast the tree into term clusters. We then selected a subset of representative terms from this cluster and convert them into a network layout(figure 6). More specifically, each term is represented by a circle node, where its size is proportional to the number of input genes fall into that term, and its color represent its cluster identity. Terms with a similarity score > 0.3 are linked by an edge[66-67]. As can be seen from figure 6, it is clear that the first polygan gather most terms and thus need a deeper analysis. The table below (Table 3) is the 10 terms in the first polygan and their related LogP value and genes. Genes HOXA9, HOXB1, FGF18, SH3PXD2B and ZMIZ1 are all included in our overlapped 16 features, which biologically support our algorithm.

| Node_No | Cluster_id | Hits | LogP |
| --- | --- | --- | --- |
| 10 | 32998381 | HOXA9\|HOXB1\|HOXC4\|FGF18\|SH3PXD2B | -5.383023217 |
| 2 | 32998368 | HOXA9\|HOXB1\|HOXC4\|TBX3\|FGF18\|SH3PXD2B | -4.804057191 |
| 3 | 32998369 | HOXA9\|HOXB1\|HOXC4\|TBX3 | -4.06992319 |
| 4 | 32998370 | HOXA9\|HOXB1\|HOXC4 | -3.815372567 |
| 5 | 32998371 | HOXA9\|HOXB1\|HOXC4\|TBX3 | -3.535152806 |
| 6 | 32998372 | HOXA9\|HOXB1\|HOXC4 | -3.440744804 |
| 7 | 32998373 | HOXA9\|HOXB1\|HOXC4\|TBX3\|TRIM15 | -3.384127137 |
| 8 | 32998374 | HOXA9\|HOXB1\|HOXC4\|TBX3 | -3.270066923 |
| 9 | 32998375 | HOXA9\|HOXB1\|HOXC4\|TBX3\|ZMIZ1 | -3.241706369 |
| 1 | 32998376 | HOXA9\|HOXB1\|HOXC4\|TBX3\|ZMIZ1 | -3.187787073 |

*Table 4 Detailed information of enriched terms in Figure 12*

### 3.5.2 Gene Annotation

In this section, we will give gene annotations to the overlapped features(intersection of 20, 30, 40, 50 solutions). The genes and corresponding relations with specific cancers are given as follows.

| IDENTIFIERS | GENE | CHROM | START | END | RELATED CANCER |
|---|---|---|---|---|---|
| CG17198308 | IFFO1 | chr12 | 6665329 | 6665331 | GBMLGG |
| CG17853216 | ZMIZ1 | chr10 | 81002479 | 81002481 | GBMLGG, PRAD |
| CG14528056 | GBAP1 | chr1 | 155194781 | 155194783 | STAD |
| CG07823492 | HOXB1 | chr17 | 46608098 | 46608100 | GBMLGG |
| CG02125316 | FGF18 | chr5 | 170878208 | 170878210 | BRCA |
| CG17295878 | TBC1D16 | chr17 | 77924664 | 77924666 | BRCA, SKCM |
| CG27009703 | HOXA9 | chr7 | 27204893 | 27204895 | LUNG |
| CG03868944 | SH3PXD2B | chr5 | 171808448 | 171808450 | PAAD |
| CG24402880 | PLAC8 | chr4 | 84035836 | 84035838 | COAD,READ |
| CG00518941 | PRKCE | chr2 | 46361963 | 46361965 | COAD,READ |
| CG13709271 | MARCH8 | chr10 | 45966141 | 45966143 | ESCA |
| CG08552167 | NKX6-2 | chr10 | 134560108 | 134560110 | HNSC |

*Table 5    12 out of 16 overlapped features have been identified related with cancer*

IFFO1-constituted nucleoskeleton prevents chromosome translocation which is a major cause of the onset and progression of diverse types of cancers by immobilizing broken DNA ends during tumorigenesis. Inactivating IFFO1 or its interaction with XRCC4 or lamin A/C leads to increases in both the mobility of broken ends and the frequency of chromosome translocation.[68]

ZMIZ1 gene encodes a member of the PIAS (protein inhibitor of activated STAT) family of proteins. The encoded protein regulates the activity of various transcription factors, including the androgen receptor, Smad3/4, and p53 [69]. Rogers(2013) demonstrated that the ZMIZ1 might confer selective advantage to tumor cells, which indicated the casual link between ZMIZ1 and tumorigenesis. [70]

GBAP1 is a pseudogene which is defined as nonfunctional gene bearing close resemblance to the known genes at different loci [71]. Research has been conducted to uncover the epigenetic regulation, biological function, and clinical application of GBAP1 in Gastric Cancer risk [72].

HOXB1 plays a important role as an oncogene in diverse tumors as result suggests that HOXB1 gene is a tumor suppressor which is regulated by miR-3175 in Glioma.[73]

FGF18 contributes to breast cancer cells following activation of the Akt/GSK3β/β-catenin pathway, which makes FGF18 a possible candidate target for breast cancers. [74]

TBC1D16 have been found associated with tumor progression and metastasis in multiple cancer types. Hypomethylation of TBC1D16 was observed in breast cancer and metastatic tumors [75].

SH3PXD2B encodes an adapter protein that is characterized by a PX domain and four Src homology 3 domains [76]. The encoded protein is required for podosome and invadopodia formation and is involved in cell adhesion and migration of numerous cell types. It has been demonstrated that inhibitors of podosome and invadopodia formation might have utility in the treatment of vascular diseases and cancer [77].

HOXA9 which is down-regulated in lung cancer and acts as a tumor progression suppressor is closely related to the aggressive growth of lung cancer cells [78]. HOXA9 contains a homeobox domain that binds to DNA and regulates downstream gene expression [79]. In addition, it is demonstrated that exogenous up-regulation of HOXA9 inhibits tumor cell invasion and migration, inhibits the expression of zinc finger 2 (SNA12/SLUG) by inhibiting the activity of the nuclear factor (NF) -kb [80].

PRKCE is generally related to the expression of protein kinase C epsilon (PKCε). Encoded by PRKCE, PKCε is an enzyme often related to cell transformation and tumorigenesis. Under the action of PKCε, the Ras/Raf pathway is activated, resulting in the transcription of genes involved in cell proliferation and growth [81].

EPAS1 gene is responsible for the expression of HIF2α, which is a key component of hypoxia-inducible factor (HIF) associated with the development of tumors in hypoxic conditions. EPAS1 is transcriptionally regulated by DNA methylation in colorectal cancer. Research on EPAS1 have revealed that, for patients with colorectal cancer, there is significant DNA hypermethylation in the EPAS1 regulatory region, which is associated with a decrease in EPAS1 mRNA level in primary cancerous tissues [82].

NKX6Bis a murine-homeobox-containing gene localized distally on Chromosome (Chr) 710q26, a region where frequent loss of heterozygosity has been observed in many brain tumors.It is suggested that NKX6B is a possible candidate tumor suppressor gene for brain tumors, particularly for oligodendrogliomas.[83]

PLAC8 is a multi-functional protein that is highly expressed in the intestine, lung, spleen, and innate immune cells, and is involved in various diseases, including cancers, obesity, and innate immune deficiency.[84]

MARCH8 has been shown to down-regulate TNF-related apoptosis inducing ligand receptor 1 (TRAIL-R1) from the cell surface at steady state and possess the potential to provide therapeutic benefit to cancer patients [85]. Additionally, Kumar(2007)identified MARCH8 as one of the differentially expressed gene in esophageal squamous cell carcinoma (ESCC) using 19.1K cDNA microarrays [86] and following the finding Shivam(2017) analyzed its expression and clinical relevance in ESCAs and observed that silencing of MARCH8 affects proliferation, migration/invasion, colony formation potential and apoptosis of ESCC cells. [87]

## 4   Conclusions and future work

In this research, we succeed in using small number of identifiers in DNA methylation 450k to classify 30 cancer types. We solve the problem of selecting a small subsets of genes from large number of DNA 450K methylated sites, and our parsimonious model is demonstrated to be efficient, achieving an accuracy of 91% , 92.3%, 93.3% and 93.5% for 20, 30, 40 and 50 features respectively. Also, 12 out of 16 of the top features we select using the framework have been shown implicated in cancer development, which sustain our algorithm biologically. Therefore, our method may act as an auxiliary diagnoistic method to determine the actual clinicopathological status of cancer type.

On the other hand, our team also pointed out the problems of distinguishing between COAD(colon adenocarcinoma) and READ(rectum adenocarcinoma) class using the methylation value, which agrees with Li (2017). Since Li uses the RNA-seq data which is somewhat different from us, it is necessary for more research to be done into the area in the future.

**APPENDIX**

**Appendix 1 : Number of samples in each class**

| Dataset | No. of samples |
|---------|----------------|
| ACC | 80 |
| BLCA | 434 |
| BRCA | 888 |
| CESC | 312 |
| CHOL | 45 |
| COAD | 337 |
| DLBC | 48 |
| ESCA | 202 |
| GBMLGG | 685 |
| HNSC | 580 |
| KICH | 66 |
| KIRC | 480 |
| KIRP | 321 |
| LAML | 194 |
| LIHC | 429 |
| LUNG | 907 |
| MESO | 87 |
| PAAD | 195 |
| PCPG | 187 |
| PRAD | 549 |
| READ | 106 |
| SARC | 269 |
| SKCM | 476 |
| STAD | 398 |
| TGCT | 156 |
| THCA | 571 |
| THYM | 126 |
| UCEC | 478 |
| UCS | 57 |
| UVM | 80 |

## Appendix 2

| #id | gene | chrom | chromStart | chromEnd | Importance |
|---|---|---|---|---|---|
| cg17198308 | IFFO1 | chr12 | 6665107 | 6665109 | . |
| cg10706013 | PDE9A | chr21 | 44104949 | 44104951 | . |
| cg05738240 | RBM19,TRNA_Pseudo | chr12 | 114337926 | 114337928 | . |

| cg22727783 | TRIO,FAM105A | chr5 | 14441073 | 14441075 | . |
| cg14528056 | GBAP1 | chr1 | 155194781 | 155194783 | Overlapped |
| cg24402880 | PLAC8 | chr4 | 84035836 | 84035838 | Overlapped |
| cg17853216 | ZMIZ1 | chr10 | 81002479 | 81002481 | Overlapped |
| cg07823492 | HOXB1 | chr17 | 46608098 | 46608100 | Overlapped |
| cg13668207 | BC043551 | chr12 | 65904451 | 65904453 | Overlapped |
| cg17295878 | TBC1D16 | chr17 | 77924664 | 77924666 | Overlapped |
| cg23473955 | ABHD15 | chr17 | 27893406 | 27893408 | Overlapped |
| cg00518941 | PRKCE,EPAS1 | chr2 | 46361963 | 46361965 | Overlapped |
| cg08552167 | INPP5A,NKX6-2 | chr10 | 134560108 | 134560110 | Overlapped |
| cg27009703 | HOXA9 | chr7 | 27204893 | 27204895 | Overlapped |
| cg13709271 | MARCH8,40976 | chr10 | 45966141 | 45966143 | Overlapped |
| cg08875705 | IFFO1 | chr12 | 6665329 | 6665331 | Overlapped |
| cg12635790 | MIR3150B | chr8 | 96112813 | 96112815 | Overlapped |
| cg18368125 | TMED6 | chr16 | 69385826 | 69385828 | Overlapped |
| cg03868944 | SH3PXD2B | chr5 | 171808448 | 171808450 | Overlapped |
| cg02125316 | FGF18 | chr5 | 170878208 | 170878210 | Overlapped |
| cg08913523 | Mir_544 | chr8 | 126649806 | 126649808 | . |
| cg04770088 | RGS9 | chr17 | 63225019 | 63225021 | . |
| cg00363813 | IFFO1 | chr12 | 6664871 | 6664873 | . |
| cg23089272 | FSIP1 | chr15 | 39985430 | 39985432 | . |
| cg06627617 | ASAP2,ITGB1BP1 | chr2 | 9471178 | 9471180 | . |
| cg17806482 | LAMB1,U3 | chr7 | 107608006 | 107608008 | . |
| cg05327192 | KLHL35 | chr11 | 75133592 | 75133594 | . |
| cg14723977 | WIZ,AKAP8L | chr19 | 15532805 | 15532807 | . |
| cg17785786 | CPE,JA611274 | chr4 | 166414665 | 166414667 | . |
| cg18920088 | MTHFD1L,U6 | chr6 | 151346408 | 151346410 | . |
| cg27285599 | ZNF750,TBCD | chr17 | 80798022 | 80798024 | . |
| cg02989244 | MPP7 | chr10 | 28657269 | 28657271 | . |
| cg07786675 | SFN | chr1 | 27189984 | 27189986 | . |
| cg06132620 | NHSL1 | chr6 | 138820502 | 138820504 | . |
| cg09053536 | TBX3 | chr12 | 115117550 | 115117552 | . |
| cg27514336 | NEK11,NUDT16P1 | chr3 | 131068940 | 131068942 | . |
| cg01979888 | IFFO1 | chr12 | 6665423 | 6665425 | . |
| cg16783204 | AK055272 | chr16 | 89119370 | 89119372 | . |
| cg04340430 | LOC440839,AK123617 | chr2 | 113967478 | 113967480 | . |

| cg15912800 | MIR196B | chr7 | 27209196 | 27209198 | . |
| cg16953816 | VPS37B,HIP1R | chr12 | 123349951 | 123349953 | . |
| cg26654807 | ZMIZ1 | chr10 | 81002217 | 81002219 | . |
| cg10704263 | CYTH1 | chr17 | 76732904 | 76732906 | . |
| cg24425838 | LOC100132111,C2CD4D | chr1 | 151812434 | 151812436 | . |
| cg27138204 | HOXC4 | chr12 | 54446099 | 54446101 | . |
| cg12904880 | TRIM15 | chr6 | 30139831 | 30139833 | . |
| cg08145381 | KIAA0317,SNORA7 | chr14 | 75172504 | 75172506 | . |
| cg03025986 | TSSC1 | chr2 | 3292756 | 3292758 | . |
| cg13630878 | MTMR15,FAN1 | chr15 | 31215821 | 31215823 | . |
| cg22202558 | CUX1 | chr7 | 101500302 | 101500304 | . |
| cg05002406 | PDLIM3 | chr4 | 186425753 | 186425755 | . |
| cg22455450 | ZNF808 | chr19 | 53038971 | 53038973 | . |
| cg19794481 | MIR141 | chr12 | 7073239 | 7073241 | . |
| cg25599924 | ACOX3 | chr4 | 8396055 | 8396057 | . |
| cg15958289 | DCPS | chr11 | 126188993 | 126188995 | . |
| cg19483007 | WWTR1,AK309441 | chr3 | 149327650 | 149327652 | . |

**Appendix 2.** *CpGs and related genes in this study*

(Illumina Methylation450k Gene Mapping can be achieved by

https://tcga.xenahubs.net/download/probeMap/illuminaMethyl450_hg19_GPL16304_TCGAlegacy)

**Appendix 3**

| MODEL | Estimator | Classifier | Max | Min | Avg |
| --- | --- | --- | --- | --- | --- |
| 1 | Random Forest | LGB | 0.86256 | 0.82341 | 0.84389 |
| 2 | Random Forest | XGBOOST | 0.85934 | 0.81211 | 0.84461 |
| 3 | Random Forest | SVC | 0.85744 | 0.81314 | 0.83311 |
| 4 | Random Forest | KNN | 0.84292 | 0.81333 | 0.82449 |
| 5 | Random Forest | MLP | 0.8614 | 0.80903 | 0.84184 |
| 6 | Decision Tree | LGB | 0.8922 | 0.85421 | 0.87632 |
| 7 | Decision Tree | XGBOOST | 0.89425 | 0.85421 | 0.87396 |
| 8 | Decision Tree | SVC | 0.89014 | 0.83984 | 0.86421 |
| 9 | Decision Tree | KNN | 0.86051 | 0.82136 | 0.84553 |
| 10 | Decision Tree | MLP | 0.88912 | 0.83778 | 0.86134 |
| 11 | Extra Tree | LGB | 0.90862 | 0.87269 | 0.88463 |
| 12 | Extra Tree | XGBOOST | 0.89733 | 0.8768 | 0.88433 |
| 13 | Extra Tree | SVC | 0.88615 | 0.85934 | 0.87355 |
| 14 | Extra Tree | KNN | 0.88398 | 0.83265 | 0.86349 |
| 15 | Extra Tree | MLP | 0.8841 | 0.85318 | 0.87108 |
| 16 | Linear SVM | LGB | 0.89862 | 0.84805 | 0.87909 |
| 17 | Linear SVM | XGBOOST | 0.89862 | 0.84805 | 0.87909 |
| 18 | Linear SVM | SVC | 0.89322 | 0.85934 | 0.8792 |
| 19 | Linear SVM | KNN | 0.89322 | 0.85216 | 0.87068 |
| 20 | Linear SVM | MLP | 0.86448 | 0.8441 | 0.85333 |

***Appendix 3*** *Model performance of 10 fold cross-validation Accuracy*